\documentclass[10pt]{article}
\usepackage{url}
\usepackage{amsmath,amsthm,amsfonts}
\usepackage{braket}
\usepackage{bm}
\usepackage{authblk}
\usepackage[colorlinks = false]{hyperref}
\usepackage[english]{babel}
\usepackage{graphicx}
\providecommand{\keywords}[1]
{
  \small	
  \textbf{\textit{Keywords---}} #1
}

\title{Portfolio Optimisation Using the D-Wave Quantum Annealer}
\author[1]{Frank Phillipson}
\author[2]{Harshil Singh Bhatia}
\affil[1]{TNO, the Netherlands Organisation for Applied Scientific Research, The Hague, The Netherlands}
\affil[2]{Department of Computer Science and Engineering, Indian Institute of Technology, Jodhpur, India}

\begin{document}

\maketitle

\begin{abstract}
    The first quantum computers are expected to perform well at quadratic optimisation problems. In this paper a quadratic problem in finance is taken, the Portfolio Optimisation problem. Here, a set of assets is chosen for investment, such that the total risk is minimised, a minimum return is realised and a budget constraint is met. This problem is solved for several instances in two main indices, the Nikkei225 and the S\&P500 index, using the state-of-the-art implementation of D-Wave's quantum annealer and its hybrid solvers. The results are benchmarked against  conventional, state-of-the-art, commercially available tooling. Results show that for problems of the size of the used instances, the D-Wave solution, in its current, still limited size, comes already close to the performance of commercial solvers.
\end{abstract}

\keywords{Quantum Portfolio Optimisation, Quadratic unconstrained binary Optimisation, Quantum annealing, Genetic algorithm}

\section{Introduction}
Portfolio management is the problem of selecting assets (bonds, stocks, commodities) or projects in an optimal way. Classical portfolio management, as introduced by Markowitz, focuses on efficient (expected) mean-variance combinations \cite{markowitz2009harry}, and has led to a broad spectrum of optimisation problems: single and multi-objective \cite{radulescu2018multi,xidonas2017robust}, single and multi-period \cite{skaf2009multi,liagkouras2018multi}, without and with \cite{fulga2008single,liagkouras2018multi} transaction costs, deterministic or stochastic \cite{pang2017stochastic,pang2019portfolio} in all possible combinations. One of the basic problems is the single objective, maximising expected return, under budget and risk constraints. Risk is expressed by the covariance matrix of all assets. In this paper we consider a variant minimising the risk, under budget and return constraints. This is applied in family trust and pension funds, where a specific return is needed for future liabilities and a low risk is desirable. This leads to a quadratic optimisation problem with continuous or binary variables. The variables are continuous if a fraction of the budget is allocated to an asset. Binary variables can be found when the choice is whether or not to invest in a specific asset or project. 

Solving this kind of problems is not trivial. Integer quadratic programming (IQP) problems are NP-hard, the decision version of IQP is NP-complete \cite{del2017mixed}. Binary quadratic programming problems are also NP-hard in general \cite{garey1979computers}, however, specific cases are polynomially solvable \cite{li2010polynomially}. Classical solvers such as CPLEX, Gurobi, Localsolver, are still getting better in solving these problems for bigger instances. Next to these solvers, heuristic approaches exist, based on meta-heuristics like Particle Swarms, Genetic Algorithms, Ant Colony and Simulated Annealing. A recent overview of these approaches for Portfolio Optimisation can be found in \cite{zanjirdar2020overview}.

Quadratic optimisation problems with binary decision variables are expected to be a sweet-spot for near future quantum computing \cite{ronagh2016solving}, using Quantum Annealing \cite{mcgeoch2014adiabatic} or the Quantum Approximate Optimisation Algorithm (QAOA) on a gate model quantum computer \cite{farhi2016quantum}. Quantum computing is the technique of using quantum mechanical phenomena such as superposition, entanglement and interference for doing computational operations. The type of devices which are capable of doing such quantum operations are still being actively developed and are called quantum computers. We distinguish between two paradigms of quantum computing devices: gate-based and quantum annealers. A practically usable quantum computer is expected to be developed in the next few years. In less than ten years quantum computers are expected to outperform everyday computers, leading to breakthroughs in artificial intelligence \cite{neumann2019machine}, the discovery of new pharmaceuticals and beyond \cite{moller2017impact,gemeinhardt2020quantum}. Currently, various parties, such as Google, IBM, Intel, Rigetti, QuTech, D-Wave and IonQ, are developing quantum chips, which are the basis of the quantum computer \cite{resch2019quantum}. The size of these computers is limited, with the state-of-the-art being around 70 qubits for gate-based quantum computers and 5000 qubits for quantum annealers. In the meantime, progress is being made on algorithms that can be executed on those quantum computers and on the software (stack) to enable the execution of quantum algorithms on quantum hardware \cite{piattini2020talavera}. 

This also means that Portfolio Optimisation is one of the promising applications of quantum computing in finance \cite{orus2019quantum}. The work of \cite{elsokkary2017financial} presents an implementation of Markowitz's portfolio selection on D-Wave, where the expected return is maximised whilst minimising the covariance (risk) of the portfolio under a budget constraint. They formulate this as an Ising problem and solve it on the D-Wave One, having 128 qubits, for 63 potential investments within 20 $\mu$s on the quantum processor. They indicate that the solution depends on the weights added to each of the objectives and constraints. The same is done in \cite{venturelli2019reverse}, where reverse quantum annealing is used to optimise the risk-adjusted returns by the use of the metrics of Sharpe ratio. In \cite{marzec2016portfolio} the stock returns, variances and co-variances are modelled in the graph-theoretic maximum independent set (MIS) and weighted maximum independent set (WMIS) structures under combinatorial optimisation. These structures are mapped into the Ising physics model representation of the underlying D-Wave One system. This is benchmarked against the MATLAB standard function quadprog.

A newer version of the D-Wave hardware is used in \cite{cohen2020portfolio}. They also perform a stock selection out of a universe of U.S. listed, liquid equities based on the Markowitz formulation and the Sharpe ratio. They approach this ﬁrst classically, then by an approach also using the D-Wave 2000Q. The results show that practitioners can use a D-Wave system to select attractive portfolios out of 40 U.S. liquid equities. The research has been extended to 60 U.S liquid equities \cite{cohen2020portfolio60}. 

Algorithms have also been created for the gated quantum computer. In \cite{rebentrost2018quantum} an algorithm is given for Portfolio Optimisation that runs in poly($log(N)$), where N is the size of the historical return data set. The number of required qubits here is also $N$. An alternative method for a gated quantum computer is given by \cite{kerenidis2019quantum}. In \cite{QuanSim} a quantum-enhanced simulation algorithm is used to approximate a computationally expensive objective function. Quantum Amplitude Estimation (QAE) provides a quadratic speed-up over classical Monte Carlo simulation. Combining QAE with quantum optimisation can be used for discrete optimisation problems like Portfolio Optimisation.

With the introduction of the 5000 qubit Advantage quantum system by D-Wave and the new hybrid solver tooling a new heuristic approach is operational for in-production quantum computing applications\footnote{https://www.D-Wavesys.com/press-releases/d-wave-announces-general-availability-first-quantum-computer-built-business}. In this paper we use the new functionality of D-Wave for the portfolio management problem and compare its performance with other state-of-the-art, commercially available tooling. 

In the remainder of this paper first the Portfolio Optimisation problem is formulated. This is a quadratic, constrained, binary optimisation problem that can in principle be solved by commercial solvers. The solvers we use in this paper as benchmark for the quantum approach are presented in Section \ref{sec:bench}. In Section \ref{sec:imple} the implementation of the problem on the D-Wave quantum annealer is shown. The results of solving instances of two main stock indices, the Nikkei225 and the S\&P500, on the quantum annealer and the comparison with the benchmark approaches, will be presented in Section \ref{sec:resul}. We end with some conclusions and recommendations. 

\section{Problem description}
We look at a Portfolio Optimisation problem, where we have $N$ assets to invest in,  $P_1,...,P_N$. The expected return of assets $i$ equals $\mu_i$ and the risk of the asset, denoted by the standard deviation, equals $\sigma_i$. The returns of the assets are correlated, expressed by the correlation $\rho_{ij}$ for the correlation between assets $i$ and $j$. We have now the return vector $\mu=\{\mu_i\}$ and the risk matrix $\Sigma=\{\sigma_{ij}\}$ where $\sigma_{ij}=\sigma_i^2$ if $i=j$ and $\sigma_{ij}=\rho_{ij} \sigma_i \sigma_j$ if $i\neq j$.

Assume that we have a budget to select $n$ assets out of $N$. We want the return to be higher than a certain value $R^*$ and are searching for that set of $n$ assets that realise the target return against minimal risk. We therefore define $x_i=1$ if asset $i$ is selected and  $x_i=0$ otherwise. This gives the following optimisation problem:
\begin{align}
    \min x^T \Sigma x, \\
    \text{s.t.} \sum_{i=1}^N x_i=n,\\
    \mu^Tx \geq R^*,
\end{align}
which is a quadratic, constrained, binary optimisation problem.

\section{Benchmarks} \label{sec:bench}
We want to compare the performance of the D-Wave hardware with other state of the art, commercially available tooling. For this we selected two solvers and two meta-heuristics.

The first solver is LocalSolver \cite{localsolver}, which is a black-box local-search solver of 0-1 programming with non-linear constraints and objectives. A local-search heuristic is designed according to the following three layers: Search Strategy, Moves and Evolution Machinery. LocalSolver performs structured Moves tending to maintain the feasibility of solutions at each iteration whose evaluation is accelerated by exploiting invariants induced by the structure of the model. Unlike other math optimisation software, LocalSolver hybridises different optimisation techniques dynamically. We used an academic licensed local version.

The second solver is Gurobi \cite{gurobi}, from Gurobi Optimisation, Inc., which is a powerful optimiser designed to run in multi-core with capability of running in parallel mode. Gurobi uses the Branch and Bound Algorithm to solve Mixed-Integer Programming (MIP) models. It is based on four basic principles: pre-solve, cutting planes, heuristics, and parallelism. Each node in the Branch and Bound search tree is a new MIP. For our current MIQP (Mixed Integer Quadratic Programming), a Simplex Algorithm is used to solve the root node. We used an academic licensed local version of Gurobi. 

The third benchmark is a standard MATLAB implementation of Genetic Algorithms, which is a method for solving both constrained and unconstrained optimisation problems based on a natural selection process that mimics biological evolution. The algorithm repeatedly modifies a population of individual solutions. It is a stochastic, population-based algorithm that searches randomly by mutation and crossover among population members.

The last benchmark is the Simulated Annealing approach as implemented by D-Wave in their Ocean environment. Their sampler implements the simulated annealing algorithm, based on the technique of cooling metal from a high temperature to improve its structure (annealing). This algorithm often finds good solutions to hard optimisation problems. 

\section{Implementation} \label{sec:imple}
In this paper we use the newly introduced (2020) functionality of D-Wave for the portfolio management problem and compare its performance with the commercial solvers described in Section 3. In this section we describe how the problem can be implemented on the D-Wave hardware. For this, the QUBO representation will be derived, a method to find the parameters is given and D-Wave's quantum and hybrid algorithms are explained.

\subsection{QUBO representation}
The D-Wave hardware solves Ising or QUBO problems. The QUBO  \cite{glover2019tutorial} is expressed by the optimisation problem:
\begin{equation} \label{eq:1}
\text{QUBO: min/max } y=x^t Qx,
\end{equation}
where $x \in \{0,1\}^n$ are the decision variables and $Q$ is a $n \times n$ coefficient matrix. Another formulation of the problem, often used, equals 
\begin{equation} \label{eq:2}
\text{QUBO: min/max } H=x^tq + x^tQx,
\end{equation}
or a combination of multiple of these terms 
\begin{equation} \label{eq:3}
\text{QUBO: min/max } H=\lambda_1 \cdot H_1 + \lambda_2 \cdot H_2 + \cdots,
\end{equation} 
where $\lambda_1, \lambda_2, \dots$ are weights that can used to tune the problem and include constraints into the QUBO. For already a large number of combinatorial optimisation problems the QUBO representation is known \cite{glover2019tutorial,lucas2014ising}. Many constrained binary programming problems can be transformed easily to a QUBO representation. Assume that we have the problem 
\begin{equation} 
\text{min } y=c^t x, \text{ subject to } Ax=b,
\end{equation}
then we can bring the constraints to the objective value, using a penalty factor $\lambda$ for a quadratic penalty:
\begin{equation} 
\text{min } y=c^t x + \lambda (Ax-b)^t(Ax-b).
\end{equation}
Using $P=Ic$, the matrix with the values of vector $c$ on its diagonal, we get
\begin{equation} 
\text{min } y=x^t Px + \lambda (Ax-b)^t(Ax-b)=x^t Px + x^t Rx + d = x^t Qx,
\end{equation}
where matrix $R$ and the constant $d$ follow from the matrix multiplication and the constant $d$ can be neglected, as it does not influence the optimisation problem.

A QUBO problem can be easily translated into a corresponding Ising problem of $N$ variables $s_i$ $(i=1,..,N)$ with $s_i\in\{\ -1,1\}$ given by :

\begin{align}
\text{min } y = \sum_{i=1}^N h_i s_i + \sum_{i=1}^N\sum_{j=i+1}^N J_{ij} s_i s_j 
\end{align}
The Ising model and QUBO model are related by $s_i = 2x_i - 1$.

\subsection{QUBO formulation}
We now create a QUBO formulation for the problem given in Equation (1)-(3). In case Equation (3) is an equality, this problem can be translated easily to a QUBO formulation:

\begin{align}
 \min \big(\lambda_0 x^T \Sigma x + \lambda_1 (\sum_{i=1}^N x_i-n)^2 + \lambda_2 (\mu^Tx - R^*)^2 \big).
\end{align}

In the case of inequalities in Equation (3) we have to add additional $K$ slack variables $y_k$ $(k=1,\dots,K)$, where $K=\lfloor \log_2(\sum_{i=1}^N (\mu_i)) \rfloor$. Scaling the $\mu$ values (in thousands, millions, etc.) will help reduce the number of variables. This leads to 

\begin{align}
 \min \big(\lambda_0 x^T \Sigma x + \lambda_1 (\sum_{i=1}^N x_i-n)^2 + \lambda_2 (\mu^Tx - R^* - \sum_{k=1}^K 2^k y_k)^2 \big).
\end{align}

\begin{figure}
    \centering
    \includegraphics[width=10cm]{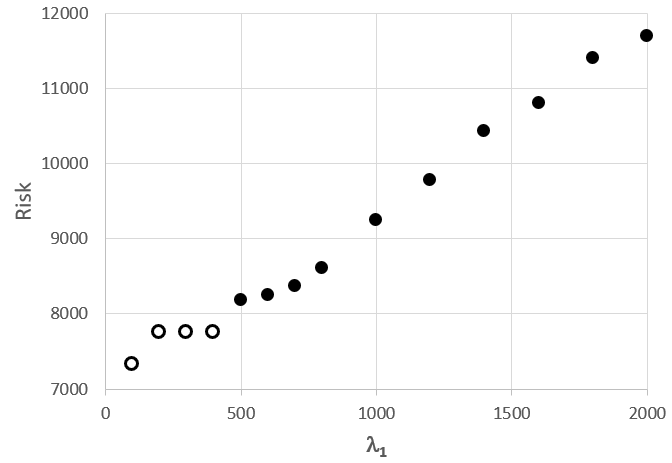}
    \caption{Relation between Risk and setting of $\lambda$. Open dots are violating the original constraints, while closed dots are valid solutions.}
    \label{fig:lambdaopt}
\end{figure}

\subsection{Finding QUBO parameters}
It is known that the performance of the D-Wave hardware is depending strongly on the choice of the penalty coefficient $\lambda$. When determining the penalty coefficients, we can set $\lambda_0=1$ and look for good values for $\lambda_1$ and $\lambda_2$. In Figure \ref{fig:lambdaopt} it is shown that the choice of $\lambda_1$ greatly influences the best found solution. Values for $\lambda_1$ lower that $500$ gives an invalid solution, i.e. the solutions do not meet the constraints given in the original problem. Values for $\lambda_1$ higher than $500$ give allowed solutions. A value of $500$ or just above would be optimal. Rule of thumb is that the gain of violating a constraint must be lower than the costs. For $\lambda_1$ this means that violating the associated constraint in the optimal solution $x^*$ for stock $i$ gives a benefit of $\sum_{j=1}^N \sigma_{ij} x_j^*$. Around the optimal value, the biggest benefit would be for stock $i$ for which the sum of the smallest $n$ values of $\sigma_{ij}$ is the largest, meaning
$$\widehat{\lambda_1}=\max_i \sum_{j=1}^n \sigma_{i\{j\}} $$
where $\sigma_{i\{j\}}$ represents the $j$-th smallest covariance value for asset $i$.

This is more complicated for $\lambda_2$. Again, we have to look at the optimal solution $x^*$ and exchange a zero and one in this solution, such that  $n$ stocks are chosen, leading to $x'$. This is done such that $(x^*-x')^T \Sigma(x^*-x') / \mu^T(x^*-x')$ is maximised. The procedure we used is as follows: 
\begin{enumerate}
    \item $A1=$ average difference between smallest $n$ sums $S_i=\sum_{j=1}^n \sigma_{i\{j\}}$,
    \item $A2=$ average positive difference in $\mu_i$ between these $n$ stocks,
    \item $\widehat{\lambda_2}=A1/A2$.
\end{enumerate}
Note that these values are a first estimation and starting point for an eventual grid search of the optimal parameters..

\subsection{Solving QUBO}
The most recent hardware, D-Wave Advantage, features a qubit connectivity based on Pegasus topology. The QUBO problem has to be transformed to this structure. Due to the current limitation of the chip size, a compact formulation of the QUBO and an efficient mapping to the graph is required. This problem is known as Minor Embedding. Minor Embedding is NP-Hard and can be handled by D-Wave's System automatically, hence we do not attempt to optimise it. While some qubits in the chip are connected using external couplers, the D-Wave QPU (Quantum Processing Unit) is not fully connected. Hence a problem variable has to be duplicated to multiple connected qubits. Those qubits should have the same
value, meaning the weight of their connection should be such that it holds in the optimisation process. All these qubits representing the same variables are part of a so-called chain, and their edge weights is called the chain strength ($\gamma$) , which is an important value in the optimisation process. In \cite{Coffrin2019ChallengesWC}, it has been indicated that if $\gamma$ is large enough, the optimal solutions will match $\gamma \geq \Sigma_{ij}|{Q_{ij}}|$. However the goal is to find the smallest value to avoid rescaling the problem. The problem of finding the smallest $\gamma$ is NP-Hard. 
On a higher level, D-Wave offers hybrid solvers. These solvers implement state of the art classical algorithms with intelligent allocation of the QPU to parts of the problem where it benefits most. These solvers are designed to accommodate even very large problems. This means that also most parameters of the embedding are set automatically. By default, samples are iterated over four parallel solvers. The top branch implements a classical Tabu Search that runs on the entire problem until interrupted by another branch completing. The other three branches use different decomposers to sample out a part of the current sample set to different samplers. The D-Wave's System sampler uses the energy impact as the criteria for selection of variables.

\section{Results} \label{sec:resul}
We solved the Portfolio Optimisation problem for two indices, the Nikkei225 and the S\&P500. We solved several instances for both indices on the D-Wave annealer, using the implementation described in Section 4, and all four benchmark approaches described in Section 3. The instances were run on a Intel(R) Core(TM) i7-8565U CPU@1.80Ghz 8GB RAM personal computer. We used the newest version of the D-Wave Leap environment solvers, Hybrid Binary Quadratic Model Version 2, for binary problems. The solver is, due to the underlying QPU, a stochastic solver. For this, we ran the hybrid solver five times for each problem instance and performed a parameter grid search for $\lambda_1$ and $\lambda_2$ for each problem instance.

We used the implementation first to select a number of stocks of the Nikkei225 index. The Nikkei225 is a stock market index for the Tokyo Stock Exchange. It has been calculated daily by the Nihon Keizai Shimbun (The Nikkei) newspaper since 1950. It is a price-weighted index, operating in the Japanese Yen, and its components are reviewed once a year. The Nikkei measures the performance of 225 large, publicly owned companies in Japan from a wide array of industry sectors. We took quarterly data of the Nikkei225 index of the last five year. The used return ($\mu$) of each stock is the five year return, and the covariance $\sigma_{ij}$ the calculated covariance over all quarters of the last 5 years. From this index, we took a subset $N$ of which $n$ are selected to minimise the risk under a budget and return ($R^*$) constraint. Note that ($R^*$) and $n$ are related here: $R^*=3000$ and $n=20$ means an average 5-year return of the portfolio of $3000/20=150$ percent.

In Table \ref{tab:paranik} we show the 20 test instances we created on the Nikkei225, each having a different combination of $(N,n,R^*)$. For each instance we show the size of the coefficient  matrix of the QUBO (Size$(Q)$), the calculated ($\lambda$) and optimal $\widehat{\lambda_2}$ value of the parameters and the resulting value of the objective function of the best solution found by the hybrid solver. Also the best known solution is shown.

\begin{table}[]
    \centering
\begin{tabular}{c c c c | c c c c | r r}
    $N$ & $n$ & $R^*$ & Size$(Q)$ & $\widehat{\lambda_1}$ & $\widehat{\lambda_2}$ & $\lambda_1$ & $\lambda_2$ & Solution & Best \\ \hline
    50 & 10 & 0 & 60 & 76 & 0 & 70 & 0 & 256 & $256^*$ \\
    50 & 25 & 0 & 60 & 320 & 0 & 365 & 0 & 3,035 & $3,035^*$ \\
    50 & 10 & 1,200 & 60 & 76 & 0.04 & 76 & 0.1 & 321 & $321^*$ \\
    50 & 25 & 1,200 & 60 & 320 & 0.10 & 365 & 0.1 & 3,058 & $3,058^*$ \\
    100 & 20 & 0 & 111 & 137 & 0 & 100 & 0 & 809 &  $809^*$\\
    100 & 50 & 0 & 111 & 607 & 0 & 700 & 0 & 10,999 & $10,999^*$\\
    100 & 20 & 2,500 & 111 & 137 & 0.05 & 100 & 0.1 & 1,126 & $1,126^*$\\
    100 & 50 & 2,500 & 111 & 607 & 0.09 & 700 & 0.1 & 11,065 & $11,065^*$\\
    150 & 20 & 0 & 161 & 116 & 0 & 100 & 0 & 628 & $628^*$\\
    150 & 50 & 0 & 161 & 502 & 0 & 500 & 0 & 7,993 & $7,993^*$\\
    150 & 20 & 3,000 & 161 & 116 & 0.04 & 100 & 0.1 & 1,271 & $1,256^*$\\
    150 & 50 & 3,000 & 161 & 502 & 0.10 & 500 & 0.1 & 8,315 & $8,270^*$\\
    200 & 20 & 0 & 211 & 100 & 0 & 100 & 0 & 530 & 529\\
    200 & 50 & 0 & 211 & 428 & 0 & 500 & 0 & 6,000 & 5,950\\
    200 & 20 & 3,250 & 211 & 100 & 0.04 & 100 & 0.1 & 1,308 & 1,262\\
    200 & 50 & 3,250 & 211 & 428 & 0.08 & 500 & 0.1 & 6,859 & 6,500\\
    225 & 20 & 0 & 236 & 88 & 0 & 88 & 0 & 484 & 482\\
    225 & 50 & 0 & 236 & 385 & 0 & 385 & 0 & 5,468 & 5,379\\
    225 & 20 & 3,500 & 236 & 88 & 0.05 & 100 & 0.1 & 1,504 & 1,455\\
    225 & 50 & 3,500 & 236 & 385 & 0.07 & 385 & 0.1 & 6,610 & 6,131\\
    
    \hline 
\end{tabular}
    \caption{20 instances for the Nikkei225 index with their parameters and HQPU solutions.}
    \label{tab:paranik}
\end{table}

In Table \ref{tab:allnik} the results of all benchmark tooling per test instance. In the table are shown the results of the hybrid solver (HQPU), Simulated Annealing (SA), Genetic Algorithm (GA), the Gurobi solver (GB) and LocalSolver (LS). We see that LocalSolver finds for all instances the best solution. Gurobi also finds this solution, however, it runs out of memory (locally) for instances bigger than $N=150$. For the instances it solves, it proves optimality by closing the optimality gap to $0\%$. LocalSolver is not able to close this optimality gap within reasonable time for most larger instances. The HQPU gives reasonable results, optimal for the smaller instances and well within $5\%$ of the optimal solution for the larger cases, with exception for the last instance.
The Simulated Annealing stays close the HQPU solution and the Genetic Algorithm implementation underperforms in comparison with the other solvers.

\begin{table}[]
    \centering
\begin{tabular}{c c c |r r r r r | r }
    $N$ & $n$ & $R^*$ & HQPU & SA & GA & GB & LS & Best Solution \\ \hline
    50 & 10 & 0 & 256 & 256 & 256 & 256 & 256 & $256^*$ \\
    50 & 25 & 0 & 3,035 & 3,035 & 3,236 & 3,035 & 3,035$^\dagger$ & $3,035^*$ \\
    50 & 10 & 1,200 & 321 & 321 & 321 & 321 & 321 & $321^*$ \\
    50 & 25 & 1,200 & 3,058 & 3,058 & 3,236 & 3,058 & 3,058 & $3,058^*$ \\
    100 & 20 & 0 & 809 & 809 & 1,102 & 809 & 809$^\dagger$ &  $809^*$\\
    100 & 50 & 0 & 10,999 & 10,999 & 12,318 & 10,999 & 10,999 & $10,999^*$\\
    100 & 20 & 2,500 & 1,126 & 1,135 & 1,450 & 1,126 & 1,126 & $1,126^*$\\
    100 & 50 & 2,500 & 11,065 & 11,065 & 12,397 & 11,065 & 11,065 & $11,065^*$\\
    150 & 20 & 0 & 628 & 628 & 1,047 & 628 & 628$^\dagger$ & $628^*$\\
    150 & 50 & 0 & 7,993 & 8,321 & 10,561 & 7,993 & 7,993$^\dagger$ & $7,993^*$\\
    150 & 20 & 3,000 & 1,271 & 1,260 & 2,937 & 1,256 & 1,256$^\dagger$ & $1,256^*$\\
    150 & 50 & 3,000 & 8,315 & 8,270 & 12,659 & 8,270 & 8,270$^\dagger$ & $8,270^*$\\
    200 & 20 & 0 & 530 & 534 & 1,151 & - & 529$^\dagger$ & 529 \\
    200 & 50 & 0 & 6,000 & 6,030 & 9,397 & - & 5,950$^\dagger$ & 5,950\\
    200 & 20 & 3,250 & 1,308 & 1,312 & 2,525 & - & 1,262$^\dagger$ & 1,262\\
    200 & 50 & 3,250 & 6,859 & 6,983 & 11,804 & - & 6,500$^\dagger$ & 6,500\\
    225 & 20 & 0 & 484 & 489 & 1,143 & - & 482$^\dagger$ & 482\\
    225 & 50 & 0 & 5,468 & 5,472 & 9,438 & - & 5,379$^\dagger$ & 5,379\\
    225 & 20 & 3,500 & 1,504 & 1,754 & 4,276 & - & 1,455$^\dagger$ & 1,455\\
    225 & 50 & 3,500 & 6,610 & 6,835 & 11,802 & - & 6,131$^\dagger$ & 6,131\\
    
    \hline 
\end{tabular}
    \caption{Results of all the methods. Values marked with * are proven optimal. Values marked with $^\dagger$ for LocalSolver are not proven optimal, solver stops before GAP=0.}
    \label{tab:allnik}
\end{table}

The performance is also depending on the computation time. In Table \ref{tab:timenik} the calculation times are listed. Here only the time the solver requires are mentioned, the time to build the problem is out of scope. Best solving times are realised by Gurobi. LocalSolver performs well regarding to finding a good solution. However, for most cases the optimality gap is not closed to $0\%$ which means the solver keeps running until the maximum admitted time is achieved.
Simulated Annealing and Genetic algorithms have increasing calculation times for the instances. The HQPU is quite fast and independent from the instance size here. The times displayed in the table correspond to one call to the hybrid solver, where we used 5 calls per instance. D-Wave allows the user to set a time limit. When the user does not specify this, a minimum time limit is calculated and used. For all instances here we did not adjust the time limit.

\begin{table}[]
    \centering
\begin{tabular}{c c c | c c c c c | c }
    $N$ & $n$ & $R^*$ & HQPU & SA & GA & GB & LS  \\ \hline
    50 & 10 & 0     & 1.0 & 8   & 58 & $<$ 1 & 2 (210) \\
    50 & 25 & 0     & 1.3 & 7  & 55 & $<$ 1 & 2 ($>$600)  \\
    50 & 10 & 1,200  & 1.0 & 2   & 53 & $<$ 1 & 2 (35)  \\
    50 & 25 & 1,200  & 0.9 & 2  & 61 & $<$ 1 & 2 (20)  \\
    100 & 20 & 0    & 1.1 & 9   & 89 & $<$ 1 & 3 ($>$600) \\
    100 & 50 & 0    & 1.1 & 20 & 88 & $<$ 1 & 2 (76) \\
    100 & 20 & 2,500 & 1.0 & 9  & 82 & $<$ 1 & 2 (556) \\
    100 & 50 & 2,500 & 1.1 & 9 & 86 & $<$ 1 & 3 (43) \\
    150 & 20 & 0    & 1.3 & 37   & 128 & $<$ 1 & 2 ($>$600) \\
    150 & 50 & 0    & 1.4 & 51  & 130 & $<$ 1 & 4 ($>$600) \\
    150 & 20 & 3,000 & 1.4 & 2 & 71 & $<$ 1 & 3 ($>$600)\\
    150 & 50 & 3,000 & 1.2 & 7  & 74 & $<$ 1 & 16 ($>$600)\\
    200 & 20 & 0    & 1.2 & 26   & 157 & - & 3 ($>$600)  \\
    200 & 50 & 0    & 1.1 & 33  & 161 & - & 4 ($>$600)\\
    200 & 20 & 3,250 & 1.1 & 28  & 106 & - & 21 ($>$600)\\
    200 & 50 & 3,250 & 1.1 & 32  & 128 & - & 7 ($>$600)\\
    225 & 20 & 0    & 1.2 & 31   & 169 & - & 3 ($>$600) \\
    225 & 50 & 0    & 1.1 & 43  & 168 & - & 12 ($>$600) \\
    225 & 20 & 3,500 & 1.3 & 32  & 131 & - & 165 ($>$600)\\
    225 & 50 & 3,500 & 1.2 & 37  & 140 & - & 207 ($>$600)\\
    
    \hline 
\end{tabular}
    \caption{Pure solver times in seconds of all the methods. For LocalSolver the time to find the best binary solution is given and between brackets the time to prove optimality. The HQPU time is per query, in the analysis we used 5 queries per problem and a parameter grid search was performed prior to these runs.}
    \label{tab:timenik}
\end{table}

The second exercise we perform on the bigger S\&P500. The S\&P500 is a stock market index that measures the stock performance of 500 large companies listed on stock exchanges in the United States. It is one of the most commonly followed equity indices. The S\&P500 index is a capitalisation-weighted index and the 10 largest companies in the index account for 26\% of the market capitalisation of the index: Apple Inc., Microsoft, Amazon.com, Alphabet Inc., Facebook, Johnson \& Johnson, Berkshire Hathaway, Visa Inc., Procter \& Gamble and JPMorgan Chase. For this index we created five instances, which are depicted with the results of the solvers in Table \ref{tab:resSP} and Table \ref{tab:timeSP}.

\begin{table}[]
    \centering
\begin{tabular}{c c c c |r r r r r | r }
    $N$ & $n$ & $R^*$ & Size$(Q)$ & HQPU & SA & GA & GB & LS & Best \\ \hline
    100 & 50 & 3500 & 112 & 5518 & 5518 & 6959 & 5518 & 5518$^\dagger$ &  $5518^*$\\
    200 & 50 & 3500 & 213 & 3121 & 3141 & 5896 & - & 3121$^\dagger$ &  $3121$\\
    300 & 50 & 3500 & 314 & 2492 & 2525 & 6825 & - & 2485$^\dagger$ &  $2485$\\
    400 & 50 & 3500 & 414 & 1537 & 1677 & 6846 & - & 1504$^\dagger$ &  $1504$\\
    500 & 50 & 3500 & 515 & 1521 & 1570 & 9014 & - & 1286$^\dagger$ &  $1286$\\
    \hline 
\end{tabular}
    \caption{Results of all the methods. Values marked with * are proven optimal. Values marked with $^\dagger$ for LocalSolver are not proven optimal, solver stops before GAP=0.}
    \label{tab:resSP}
\end{table}

Again we see that Gurobi is not able to run instances bigger than $N=150$. LocalSolver finds the best solutions but is not able to close the optimality gap. In the three largest instances the gap stayed $100\%$ within the allowed 600s. Genetic algorithms performed poorly again. The hybrid approach and Simulated Annealing were able to stay close to the LocalSolver solutions, within $3\%$ except for the last instance. Although we gave the HQPU more time to solve, we adjusted the time limit by hand, it stayed on $18\%$ above the best solution found by LocalSolver, as the Simulated Annealing solver did. The solving times of the HQPU stay quite reasonable as compared to the other solvers.

\begin{table}[]
    \centering
\begin{tabular}{c c c | c c c c c | c }
    $N$ & $n$ & $R^*$ & HQPU & SA & GA & GB & LS  \\ \hline
    100 & 50 & 3500     & 1.0 & 8   & 61 & $<$ 1 & 2 ($>$600) \\
    200 & 50 & 3500     & 1.3 & 28   & 92 & - & 3 ($>$600) \\
    300 & 50 & 3500     & 1.6 & 52  & 124 & - & 7 ($>$600) \\
    400 & 50 & 3500     & 1.9 & 92  & 155& - & 18 ($>$600) \\
    500 & 50 & 3500     & 5.8 & 135   & 221 & - & 16 ($>$600) \\
    \hline 
\end{tabular}
    \caption{Pure solver times in seconds of all the methods. For LocalSolver the time to find the best binary solution is given and between brackets the time to prove optimality. The HQPU time is per query, in the analysis we used 5 queries per problem and a parameter grid search was perfumed prior to these runs.}
    \label{tab:timeSP}
\end{table}

\section{Conclusions and Further research}
Quantum computing is still in its early stage. However, the moment of quantum computers outperforming classical computers is coming closer. In this phase, hybrid solvers, using classical methods combined with quantum computing, are promising. The quantum paradigm that is already maturing is quantum annealing combined with the Hybrid Solvers that D-Wave is offering. In this paper we showed the performance of this hybrid approach, applied to a quadratic optimisation problem occurring in finance, Portfolio Optimisation. Here a portfolio of assets is selected, minimised the total risk of the portfolio. This portfolio has to meet some return and budget constraints. 

We suggested an implementation of this problem on the newest D-Wave quantum annealer using the hybrid solver. The performance on a problem where 50 stocks needs to be selected from an existing index, the Nikkei225 or the S\&P500, was already reasonable, in comparison with the performance of other commercially available solvers. Of these solvers, LocalSolver performed the best, finding the optimal solution in all cases very quickly. However, proving optimality, and thus finishing the optimisation task, was not realised within reasonable time. The hybrid quantum solver was able to find a solution within $3\%$ of the optimal solution for the S\&P500 index up to 400 stocks. Further improvement of the hybrid solvers and the enlargement of the QPU in the coming years lead to the expectation that this computing paradigm is close to real business applications.

For further research we recommend to dive into the hybrid approach and look for improvements. The used method is a standard method, where D-Wave offers the opportunity to tailor the methodology in the Leap environment. Also other quantum optimisation algorithms could be used, like the QAOA algorithm on a gated quantum computer.

\section*{Acknowledgments}
The authors want to thank Hans Groenewegen for providing the financial data underlying this research and Irina Chiscop for her valuable review and comments.

\bibliographystyle{unsrt}
\bibliography{biblio}
\end{document}